# Performance Analysis of Combine Harvester using Hybrid Model of Artificial Neural Networks Particle Swarm Optimization

Laszlo Nadai
*Kalman Kando Faculty of Electrical Engineering*
*Obuda University*
Budapest, Hungary
nadai@uni-obuda.hu

Felde Imre
*Kalman Kando Faculty of Electrical Engineering*
*Obuda University*
Budapest, Hungary
0000-0003-4126-2480

Sina Ardabili
*Institute of advanced studies*
*University of Pannonia*
Koszeg, Hungary
0000-0002-7744-7906

Tarahom Mesri Gundoshmian
*Dep of Biosystem Engineering*
*Uni. of Mohaghegh Ardabili*
Ardabil, Iran
0000-0002-7302-7269

Pinter Gergo
*John von Neumann Faculty of Informatics*
*Obuda University*
Budapest, Hungary
0000-0003-4731-3816

Amir Mosavi [1,2]*
[1] *Department of Mathematics and Informatics, J. Selye University*
Komarno, Slovakia
[2] *Bauhaus Universität Weimar*
Weimar, Germany
0000-0003-4842-0613

*Abstract*—Novel applications of artificial intelligence for tuning the parameters of industrial machines for optimal performance are emerging at a fast pace. Tuning the combine harvesters and improving the machine performance can dramatically minimize the wastes during harvesting, and it is also beneficial to machine maintenance. Literature includes several soft computing, machine learning and optimization methods that had been used to model the function of harvesters of various crops. Due to the complexity of the problem, machine learning methods had been recently proposed to predict the optimal performance with promising results. In this paper, through proposing a novel hybrid machine learning model based on artificial neural networks integrated with particle swarm optimization (ANN-PSO), the performance analysis of a common combine harvester is presented. The hybridization of machine learning methods with soft computing techniques has recently shown promising results to improve the performance of the combine harvesters. This research aims at improving the results further by providing more stable models with higher accuracy.

I. INTRODUCTION

Performance optimization of the combine harvesters is of utmost importance in the agricultural industry to obtain higher efficiency [1, 2]. The agricultural machinery industry uses various methods of design optimization to improve machine performance [3]. The machine parameters have been optimized using various mathematical optimization techniques as well as soft computing methods, e.g., fuzzy-based methods, multi-objective optimization, multiple criteria decision-making, and design of experiments (DOE) [4-10].

Recently, machine learning methods have been also used to model the problem. However, the application of machine learning has been mainly limited to artificial neural networks (ANNs) [11-15]. Gundoshmian et al. [16] compared the model accuracy of the adaptive neuro-fuzzy inference system (ANFIS) with the radial basis function (RBF), which is a single-layer artificial neural network, in improving the performance of a John Deere 1055 Combine harvester. Both models of ANFIS and RBF of neural networks have delivered promising results in improving machine performance. However, RBF provided better results. It has been reported that ANFIS is not able to model the output variables simultaneously, and every output variable requires individual training. Besides, modeling several outputs variables and studying various design parameters using ANNs would be possible.

Furthermore, the ANN model is shown to be faster and more stable. Consequently, the contribution of this paper is to advance an ANN-based model to further improve the performance of the combine harvester. To do so, the hybrid method of ANN-PSO is proposed which provides an optimized neural network with a high level of adaptation, stability, and generalization.

In section II., the data and materials are presented where the details of the combine harvester and the hybrid method are described. In section III., the results of the modeling and the comparative analysis of ANN-PSO and ANN are presented.

## II. MATERIALS AND METHODS

### A. Data

Data has been obtained from a John Deere 1055 Combine harvester in operation. Fig.1 represents a schematic illustration of the unit. The threshing drum (TD) has a diameter of 610 mm with a length of 1080, which includes 8 blades of the rotational speed of 410 to 1160 rpm. There are adjustable and modular slider concaves. The Sieves have an area of 1.2 m$^2$. The fan has a speed of 440 to 1060 rpm with five mechanical adjustable vanes. Thresher drum and concave have a distance of 10 mm in input and 3 mm in output [17,18].

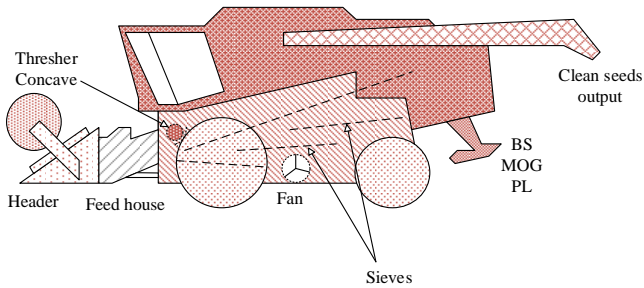

Fig. 1. Schematic representation combine harvester mechanism and variables

The data collection in operation has been performed in the form of the factorial test including three repetitions in three sets of variables. In this study, the independent variables, according to Fig. 1 are; the distance of TD and concave (A), the fan speed (B) and the Sieves openness (C). The dependent variables are; the number of broken seeds inside the tank (BS), product loss (PL), and material other than grain (MOG).

### B. Hybrid machine learning methods

The ANN can be considered as one of the most common and popular intelligent approaches for computational purposes. This method performs as a biological nervous system for prediction and estimation of target values based on input values in an undefined system without the need for systematic relationships. This method was introduced by McCulloch and Pitts [19]. ANN has been successfully applied in different fields such as engineering, agriculture, industrial problems as well as medical researches for prediction, classification, signal processing, and system modeling. ANN has three main layers including input, hidden and output layers. All layers are connected via a set of neurons. The hidden layer can include one or more sets of neurons called hidden neurons. Fig. 2 presents the schematic diagram of ANN developed in the present study.

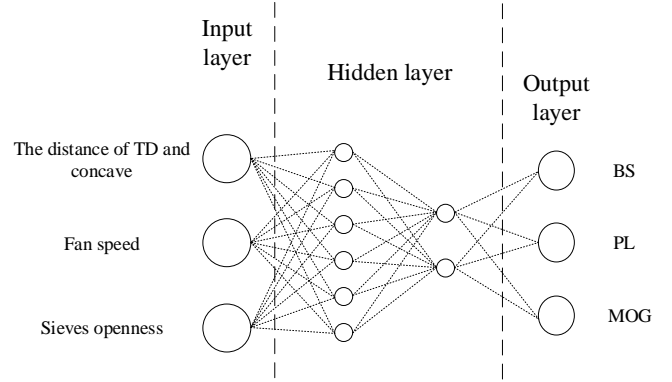

Fig. 2. Schematic representation of the ANN model and the layers

In the proposed ANN model, there exist three inputs including the distance of TD and concave, fan speed and sieves openness. The number of neurons and their sets in the hidden layer can be obtained by trial and error method in different runs. In the present study, the best arrangement of the hidden layer was six neurons in the first set connected with two neurons in the second set for generating three outputs, including BS, PL, and MOG in a combine harvester. Accordingly, the architecture of the best network was obtained to be 3-6-2-3.

The proposed mechanism of the ANN model is in a way that every neuron generates outputs based on Eq. 1 for each input ($x_j$ for $j$=1, 2, …, n) in the presence of weights [20] (for $i$=1, 2, …, n).

$$y = \varphi\left(\sum_{j=1}^{n} w_i x_j\right) \quad (1)$$

The function (1) participates in the sigmoidal function which is the most popular transfer function in the feed-forward neural network, to generate the output value (2) as follows.

$$y = \frac{1}{1+\exp\left(-\sum_{j=1}^{n} w_i x_j\right)} \quad (2)$$

Nevertheless, the ANN has some disadvantages such as long time-consuming in training process, lack of using optimal global solution, and lack of stability for network outputs in similar training situations, which directly affect its accuracy and performance in forecasting the target values. These issues made researchers use optimizers for improving the leakages of the ANN method. One of the most frequently used optimizers (PSO) was employed in the present study to improve the performance of the ANN method in developing a predictive model for the performance of combine harvester.

$$u_i(t+1) = IW u_i(t) + A_1 r_1[pbest(t) - xp_i(t)] + A_2 r_2[gbest(t) - xp_i(t)] \quad (3)$$

PSO was introduced by Kennedy and Eberhart [21] for simulating social behavior. This method considers the cooperative behavior of particles that are associated with random positions in the N-dimensional matrix. The position of particles in this space changes by the number of iterations and swarm size numbers upon the velocity updates. PSO employs Eq. 3 and 4 to update the velocity and position of each particle [4].

$$xp_i(t+1) = xp_i(t) + u_i(t+1) \quad (4)$$

Where $u$ is the velocity, IW is the initial weight, $A_1$ and $A_2$ are acceleration constant, and $x_p$ is considered as particle position. The last step of the algorithm for finishing accrues after taking the best solution. The optimum performance of this method is according to Fig 3. In the case of using ANN-PSO, the cost function depends on the weights and bias values of the ANN network [22].

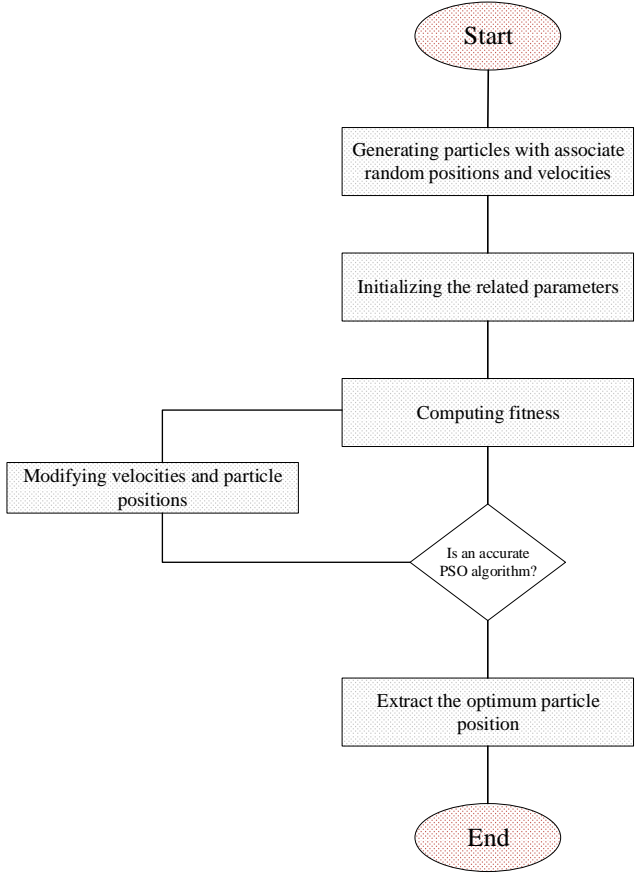

Fig. 3. Flowchart of the PSO model

## C. Evaluation metrics

The evaluation metrics of root mean square error (RMSE), Pearson correlation coefficient (R), and mean absolute error (MAE) are used to evaluate the performance of the models as follows.

In the present study, three best models were selected as the developed ANN-PSO among other runs to be compared with ANN. One with swarm size 100 at maximum iteration 186, second with swarm size 200 at maximum iteration 180 and third with swarm size 300 at maximum iteration 221.

$$RMSE = \sqrt{\frac{1}{N}\sum_{i=1}^{N}(A-P)^2} \quad (5)$$

$$R = \left(1 - \left(\frac{\sum_{i=1}^{n}(A-P)^2}{\sum_{i=1}^{n}A_i^2}\right)\right)^{1/2} \quad (6)$$

$$MAE = \frac{\sum_{i=1}^{n}|A-P|}{N} \quad (7)$$

In both stages of training and testing, the RMSE, $R^2$, and MAE values are calculated.

## III. RESULTS

The training and testing process of ANN and ANN-PSO models are performed and the results were extracted. The results of the training are given in table I. To implement the model, MOG, BS, and PL are considered as the independent variables, and the distance of TD and concave, fan speed, and Sieves openness are the independent inputs variables. Here, 70% of the data has been used for training. The model number 4, i.e., ANN-PSO, with the maximum number of iterations of 221, and PSO swarm size of 300 delivered the RSME values of 0.0371, 8.598612, and 0.385558, for BS, PL, and MOG simultaneously. The $R^2$, is also reported to be 0.977, 0.978, and 0.948 for BS, PL, and MOG, simultaneously. Consequently, model number 4 outperforms other models in terms of accuracy in training.

Furthermore, in order to train target networks, 30% of the data has been used for testing to develop the network. This stage is performed to create a precise network for the test stage. The results of training for ANN and ANN-PSO models are presented in Table. II.

TABLE I.  TRAINING RESULTS

| Model NO. | Method | Structure | RMSE | | | Correlation coefficient | | |
|---|---|---|---|---|---|---|---|---|
| | | | BS | PL | MOG | BS | PL | MOG |
| 1 | ANN | 3-6-2-3 | 0.1055 | 11.35336 | 0.648056 | 0.8572 | 0.9545 | 0.8378 |
| 2 | ANN-PSO | Max it.=186 Swarm size =100 | 0.0794 | 8.714537 | 0.502797 | 0.9 | 0.9783 | 0.908 |
| 3 | ANN-PSO | Max it.=180 Swarm size =200 | 0.0817 | 8.226392 | 0.59794 | 0.917 | 0.979 | 0.929 |
| **4** | **ANN-PSO** | **Max it.=221 Swarm size=300** | **0.0371** | **8.598612** | **0.385558** | **0.977** | **0.978** | **0.948** |

TABLE II. TESTING RESULTS

| Model NO. | Method | Structure | RMSE | | | Correlation coefficient | | |
|---|---|---|---|---|---|---|---|---|
| | | | BS | PL | MOG | BS | PL | MOG |
| 1 | ANN | 3-4-2-3 | 0.06754 | 12.1838 | 0.6718 | 0.955 | 0.899 | 0.73 |
| 2 | ANN-PSO | Max it.=186 Swarm size =100 | 0.08572 | 9.2143 | 0.4189 | 0.924 | 0.924 | 0.914 |
| 3 | ANN-PSO | Max it.=180 Swarm size =200 | 0.06061 | 8.7171 | 0.5697 | 0.981 | 0.948 | 0.927 |
| **4** | **ANN-PSO** | **Max it.=221 Swarm size=300** | **0.05239** | **6.7338** | **0.2059** | **0.97** | **0.98** | **0.993** |

In testing, the model number 4, i.e., ANN-PSO, with the maximum number of iterations of 221, and PSO swarm size of 300 delivered the RSME values of 0.05239, 6.7338, and 0.2059 for BS, PL, and MOG simultaneously. The $R^2$, is also reported to be 0.97, 0.98, and 0.993 for BS, PL, and MOG, simultaneously. Consequently, model number 4 outperforms other models in terms of accuracy in training. ANN-PSO shows better results compared to ANN. Furthermore, Fig. 4 and Fig. 5 present the predicted values with $R^2$ for BS, PL, and MOG, considering ANN and ANN-PSO, simultaneously. The comparative analysis of the deviation from the target value for all the four models is given in Fig. 6 where ANN-PSO has delivered the minimum deviation.

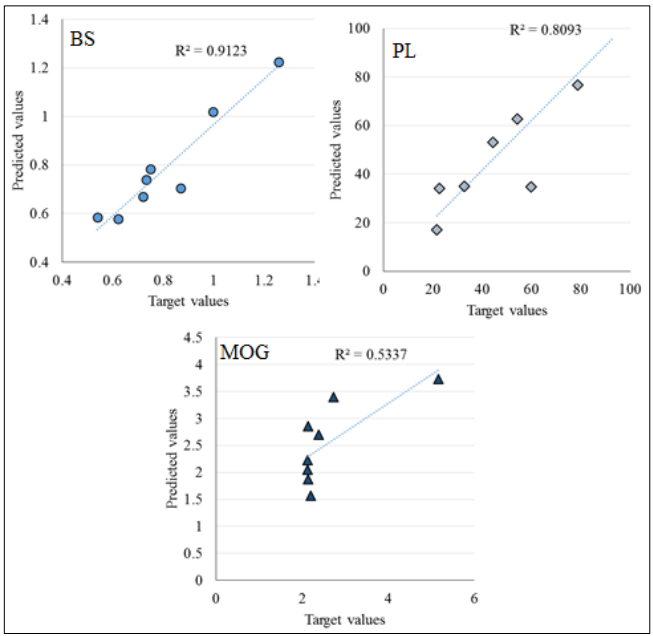

Fig. 5. Model.1: ANN model predicted values and $R^2$ for BS, PL, and MOG

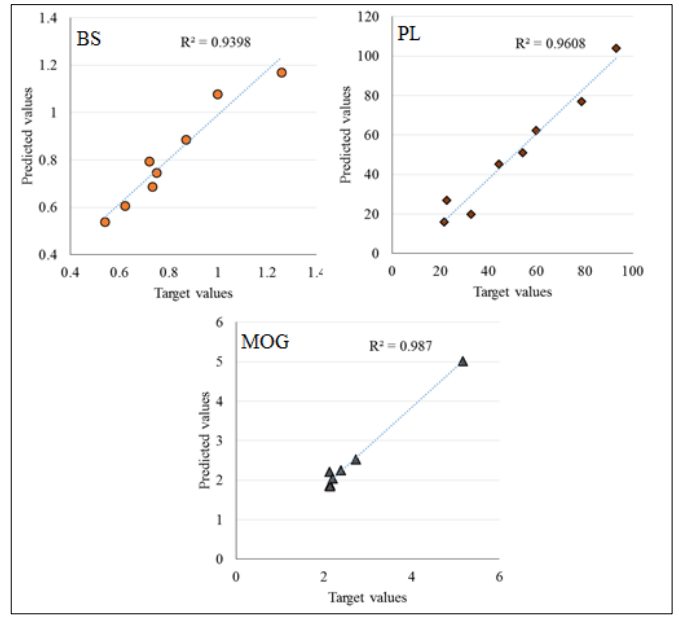

Fig. 4. Model.4: ANN-PSO model with Swarm size of 300, predicted values and $R^2$ for BS, PL, and MOG

IV. CONCLUSION

The paper presented a hybrid model of ANN-PSO for performance analysis of a common combine harvester. The results have been compared with the developed ANN model. The proposed ANN-PSO outperformed the ANN model, and the comparative analysis of the deviation from the target value reported promising. The hybrid method of ANN-PSO was successful in providing an optimized neural network with a high level of adaptation, stability, and generalization. The hybridization of machine learning methods has shown to be an essential approach to improve the performance of the prediction models. For the future research, advancement of hybrid and ensemble machine learning models, e.g., [23-28], and comparative analysis with deep learning models, e.g., [29-32] are proposed to identify models with higher efficiency.

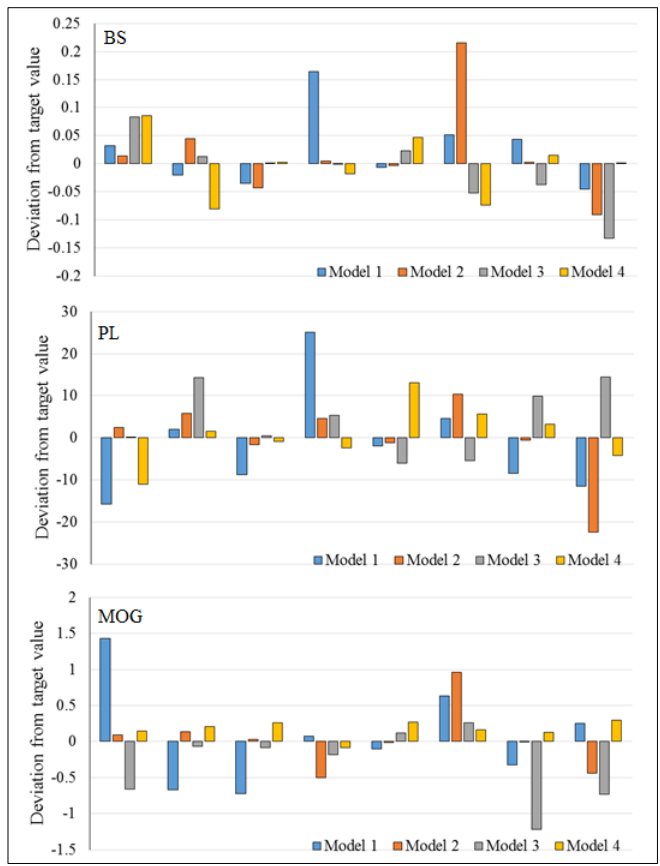

Fig. 6. The comparative analysis of the deviation from the target value for all the four models, where model 4 outperforms other models

ACKNOWLEDGMENT

We acknowledge the financial support of this work by the Hungarian State and the European Union under the EFOP-3.6.1-16-2016-00010 project and the 2017-1.3.1-VKE-2017-00025 project.